\documentclass[trackchanges]{aastex701}

\usepackage{boondox-cal}
\begin{document}

\title{QFP Waves Driven by the Tuning-Fork Effect during Magnetic Reconnecion }

\author[orcid=0000-0001-9828-1549]{Jialiang Hu}
\affiliation{State Key Laboratory of Solar Activity and Space Weather, National Astronomical observatories, Chinese Academy of Sciences, 100101,Beiing,China}
\affiliation{School of Astronomy and Space Science, University of Chinese Academy of Sciences, 100049, Beijing, China}
\email{hujl@bao.ac.com}  

\author[0000-0003-2875-7366]{Xiaozhou Zhao}
\affiliation{Yunnan Observatories, Chinese Academy of Sciences, P.O. Box 110, Kunming, Yunnan 650216, People's Republic of China}
\email{zhaoxz@ynao.ac.cn}

\author[orcid=0000-0001-9828-1549]{Guiping Zhou}
\affiliation{State Key Laboratory of Solar Activity and Space Weather, National Astronomical observatories, Chinese Academy of Sciences, 100101,Beiing,China}
\affiliation{School of Astronomy and Space Science, University of Chinese Academy of Sciences, 100049, Beijing, China}
\email{gpzhou@nao.cas.cn}  

\author[0000-0002-8077-094X]{Yuhao Chen}
\affiliation{School of Earth and Space Sciences, Peking University, Beijing 100781, People’s Republic of China}
\email{yuhao_chen@pku.edu.cn}

\correspondingauthor{Hanxian Fang, Xinping Zhou }
\email{fanghx@nudt.edu.cn,xpzhou@sicnu.edu.cn}

\author[0000-0003-4763-0854]{Chunlan Jin}
\affiliation{State Key Laboratory of Solar Activity and Space Weather, National Astronomical observatories, Chinese Academy of Sciences, 100101,Beiing,China}
\email{cljin@nao.cas.cn}

\author[0000-0002-9201-5896]{Mijie Shi}
\affiliation{Shandong Key Laboratory of Optical Astronomy and Solar–Terrestrial Environment, School of Space Science and Technology, Institute of Space Sciences, Shandong University, Weihai 264209, People’s Republic of China}
\email{shimijie@sdu.edu.cn}

\author[0009-0002-6808-5330]{Guanchong Cheng}
\affiliation{Yunnan Observatories, Chinese Academy of Sciences, P.O. Box 110, Kunming, Yunnan 650216, People's Republic of China}
\email{yhc@ynao.ac.cn}

\author[orcid=0000-0002-5499-7817,sname='']{Xiaoxia Yu}
\affiliation{State Key Laboratory of Particle Astrophysics, Institute of High Energy Physics, Chinese Academy of Sciences, Beijing 100049, China\\}
\email{xxyu@ihep.ac.cn}

\author[orcid=]{Jing Ye}    
\affiliation{Yunnan Observatories, Chinese Academy of Sciences, P.O. Box 110, Kunming, Yunnan 650216, People's Republic of China}
\email{yj@ynao.ac.cn}

\author[orcid=0000-0001-9374-4380]{Xinping Zhou}
\affiliation{College of Physics and Electronic Engineering, Sichuan Normal University, Chengdu 610068, People's Republic of China}
\email{xpzhou@sicnu.edu.cn}

\author[orcid=]{Hanxian Fang}
\affiliation{College of Advanced Interdisciplinary Studies, National University of Defense Technology, Changsha, China.}
\email{fanghx@nudt.edu.cn}

\begin{abstract}
Through three-dimensional MHD simulations, we have uncovered a kind of fast coronal wave originating from both ends of a current sheet (CS) during a solar eruption.  These waves are observed to appear near the top and bottom ends of the reconnection-related CS. The simulations demonstrate the presence of termination shock regions above the two ends of the CS. As the reconnection outflows escape from the vertical CS and encounter these termination shocks, they undergo partial reflection, redirecting towards the CS terminal fork walls. The identified waves propagate rapidly at a speed of approximately 1400 km/s with a period of just 2 s. Concurrently, the time-evolution of intensity within a small region of the CS terminal fork structures, exhibits a similar oscillation period of 2 s.  All these evidence supports the notion that these \textbf{QFP (Quasi-periodic Fast-Propagating) waves} were excited by tuning fork effects within the CS system. Essentially, the rapid reconnection outflows are reflected by the terminal shocks, striking the fork walls at the CS ends. Moreover, parts of the oscillations along the tuning fork handle are transformed into thermal energy, accumulating in the CS center and elevating the temperature. This is the first time to report such \textbf{QFP waves} resulting from tuning fork effects within the CS during a solar eruption. These waves are anticipated to manifest closely following the propagation of CMEs and adjacent to the related post-flare loops in observations, with partial confirmation in current observations.
\end{abstract}


\section{Introduction}

Magnetic reconnection plays a critical role in solar eruptions, governing both the energy release process and dynamics in the solar corona. Solar eruptions are frequently accompanied by various modes of coronal waves. The propagation characteristics of coronal waves (e,g., phase speeds, dispersion relations, and attenuation scales) encode valuable information about the local plasma-$\beta$ and magnetic topology, making them useful tools for probing the fine-scale physical processes during magnetic reconnections \citep{2022SoPh..297...20S}. Previous studies that have reported a wide range of coronal wave modes have significantly advanced our understanding  of solar activity \citep{2020ARA&A..58..441N}. However, the mechanisms by which different modes of coronal waves are generated during magnetic reconnection remain poorly understood.

Based on multi-wavelength observations, coronal waves have been found to be driven by various solar phenomena, such as flare processes, filament or flux-rope eruptions, and other related activity. For example, \citet{2022ApJ...938...24L} reported that \textbf{QFP waves}  can appear near the flare core region with a dominant period of 181 seconds, showing temporal synchronization with other quasi-periodic pulsations during the flare. \textbf{EUV imaging} observations have suggested a significant correspondence between the characteristic frequency associated with energy release  and that of the detected coronal waves during flares \cite{2012ApJ...753...53S}. \textbf{QFP waves} may be accompanied by radio burst emissions during solar eruptions  \citep{2013A&A...554A.144Y,2017ApJ...844..149K}. In addition, \citet{2016ApJ...823L..19Z} found that fast-mode coronal waves can serve as important precursors, appearing about 20 minutes before an extreme solar eruption. Moreover, \textbf{QFP waves} can also trigger secondary magnetic reconnection, leading to subsequent eruptions \citep{2020APJ...905..150Z}. Collectively, these observational results suggest a close correlation between solar eruptive activities and coronal waves. With advancements in observations and simulations, it is becoming increasing possible to uncover the generation mechanisms of the various coronal waves observed in  the solar corona.

Numerical simulations offer an effective approach to uncover the mechanisms behind coronal waves observed on the Sun. Using a 2D simulation of filament-free  flare processes,  \citet{2016ApJ...823..150T} proposed that \textbf{QFP waves} can be excited by oscillations above the loop-top during a flare. Based on a 2.5 MHD simulation, \citet{2015ApJ...800..111Y} reported that \textbf{QFP waves} can be generated by the interaction between plasmoids in the  current sheet and the surrounding magnetic structures. In a 3D MHD simulation conducted by \citet{2011ApJ...740L..33O}, it was found that \textbf{QFP waves} can be induced by quasi-periodic sources at the base of flare loops. \textbf{It is important to note that an alternative interpretation for the observed quasi-periodic patterns exists, which does not necessarily require a periodic energy source. This scenario attributes the formation of QFP waves to the wave dispersion caused by the perpendicular structuring of the coronal plasma, see, (e.g., \cite{2005SSRv..121..115N}, and \cite{2013A&A...550A...1K}), including a recent 3D simulation (e.g., \cite{2025ApJ...990....1S}.)}

In the context of a filament system, \citet{2024RAA....24l5011H} employed a 2D simulation to effectively explain \textbf{QFP waves} observed at the flank of a coronal mass ejection (CME). These waves was suggested to originate from perturbations within the filament. Subsequent 3D simulations \citep{2024RAA....24l5011H} revealed that these waves exhibit a dome-like geometry and displayed three distinct components. Understanding how these diverse coronal waves are generated during magnetic reconnection could shed light on the intricate physical processes of solar eruptions , yet this remains a crucial unresolved issue. 

Through 3D simulations, this study reveals a kind of fast coronal wave produced near the ends of the reconnection current sheet (CS). For the first time , we demonstrate that this fast coronal wave is generated by the tuning-fork effect at both ends of  the CS, and that it can appear both in the aftermath of a propagating CME and near the tops of post-flare loops.Section \ref{sec:setup} presents detailed numerical methods, while Section \ref{sec:result} provides the analysis of wave characteristics and excitation mechanisms, followed by discussions and a summary in  Section \ref{sec:sum} .
 
\section{Setup in the Numerical  Simulation} \label{sec:setup}
To study wave generation mechanisms during solar filament eruptions, we perform 3D MHD simulations using the MPI-AMRVAC code \citep{2012JCoPh.231..718K,2014ApJS..214....4P,2018ApJS..234...30X}.Our investigation focuses on plasma dynamics by solving the full set of MHD equations incorporating both gravitational effects and anisotropic thermal conduction. The governing equations are expressed as follows:
\begin{equation}
\frac{\partial \rho}{\partial t} + \nabla \cdot (\rho \mathbf{v}) = 0 \label{mass_conservation}
\end{equation}
\begin{equation}
\frac{\partial e}{\partial t} + \nabla \cdot \left[(e + P^*)\mathbf{v} - (\mathbf{v} \cdot \mathbf{B})\mathbf{B}\right] = \rho \mathbf{g} \cdot \mathbf{v} + \nabla \cdot \left(\eta \mathbf{B} \times (\nabla \times \mathbf{B}) - \mathbf{F}_{\mathrm{c}}\right) \label{energy_evolution}
\end{equation}
\begin{equation}
\frac{\partial (\rho \mathbf{v})}{\partial t} + \nabla \cdot \left[\rho \mathbf{v} \mathbf{v} + P^* \mathbf{I} - \mathbf{B} \mathbf{B}\right] = \rho \mathbf{g} \label{momentum_conservation}
\end{equation}
\begin{equation}
\frac{\partial \mathbf{B}}{\partial t} = \nabla \times \left(\mathbf{v} \times \mathbf{B} - \eta \nabla \times \mathbf{B}\right) \label{induction_equation}
\end{equation}
\begin{equation}
P^{*} = p + \frac{1}{2}|\mathbf{B}|^{2} \label{total_pressure}
\end{equation}
\begin{equation}
p = \rho T \label{ideal_gas}
\end{equation}
where $\rho$ denotes mass density, $\mathbf{v}$ flow velocity, $\mathbf{B}$ magnetic field, p gas pressure,  and T temperature. Thermal conduction follows the Spitzer mode with $ \textbf{F}{c}=-\kappa_{\parallel}( \nabla T \cdot \hat{\textbf{B}})\hat{\textbf{B}}-\kappa_{\perp}( \nabla T-( \nabla T\cdot\hat{\textbf{B}})\hat{\textbf{B}})$ where $\kappa_{\parallel}$ and  $\kappa_{\perp}$ are parallel and perpendicular conduction coefficients. The total energy density $e$ comprises kinetic, thermal, and magnetic components:  $e = \rho v^{2} /2+ p/(\gamma-1)+B^{2}/2$. All physical quantities are normalized by using characteristic scales: length $L_{0}=5\times 10^{9}$~cm, number density $n_{0} = 10^{10}$~cm$^{-3}$, and temperature $T_{0}= 10^{6}$~K. From these, we derive the characteristic density $\rho_{0} =1.673\times10^{-14}$~g~cm$^{-3}$, magnetic field strength $B_{0}=5.89$~G, plasma pressure  $P_{0} =2.76$ dyne ~cm$^{-2}$, and velocity $v_{0}=128.5$~km~s$^{-1}$. The simulation domain spans 8~L$_{0}$ in each direction, discretized using a uniform grid of $2000^{3}$ cells. The model utilizes a uniform resistivity of $\eta = 1.0\times10^5$ in normalized units. It is important to note that this constant resistivity sets a fixed and globally uniform reconnection rate. This approach does not capture the localized enhancement of resistivity expected in anomalous effects. We employ a Godunov-type finite volume scheme, utilizing the Harten–Lax–van Leer (HLL) approximate Riemann solver and a third-order Runge–Kutta time integration method. The boundary conditions consist of a line-tied lower boundary and open upper/lateral boundaries to permit plasma outflow. For further details on the numerical setup, please refer to \cite{2024RAA....24l5011H} and \cite{2023ApJ...955...88Y}. We note that the governing equations employed in this model do not include the effects of heating and radiative cooling . Thus, our model does not capture the physics of wave-induced thermal misbalance, which has been proposed as an alternative mechanism for QFP wave formation (e.g., \cite{2019PhPl...26h2113Z} ).

\begin{figure*}[ht!]
\plotone{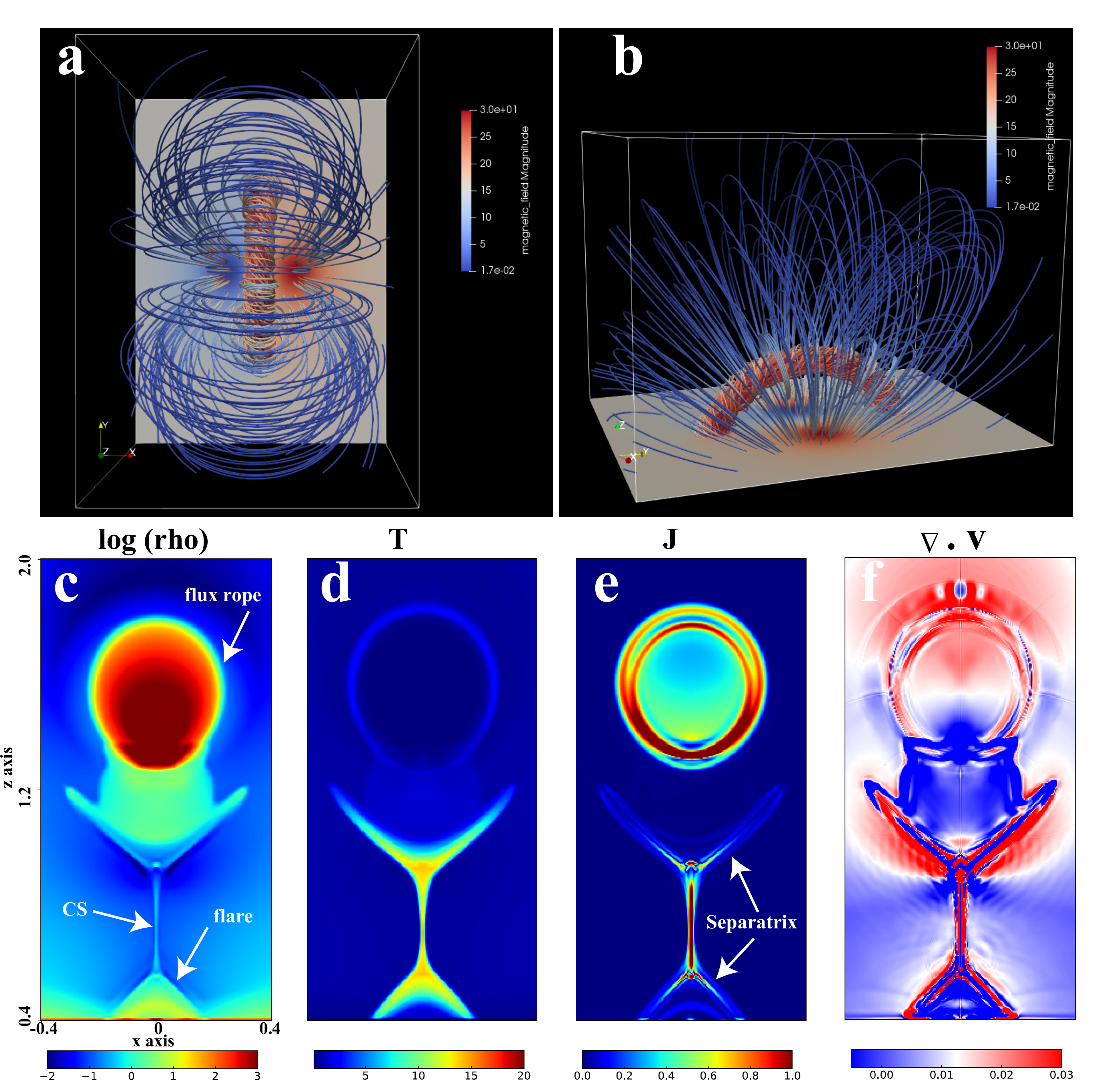}
\caption{The distributions of initial magnetic field at t = 0 s and multiple physical parameters at t = 133.09 s. \textbf{(a)-(b)} shows the initial magnetic field configuration from two perspectives. This model incorporates a flux rope with axial current $I$, a subphotospheric charge pair ($\pm q_{1}$, separation L, depth $d_{1}$) and a buried magnetic dipole ($q_{2}$, depth $d_{2}$). This configuration reproduces characteristic active region topology with $B \propto z^{-2}$ field decay.  The bottom shows spatial distributions of multiple physical parameters in the solar  atmoshpere at t = 133.09 s, including \textbf{(c)} logarithmic density, \textbf{(d)} temperature , \textbf{(e)} current density and \textbf{(f)} velocity divergence.  The current sheet (CS) is indicated, and ``Separatrix" denotes the boundary between the CS and the ambient coronal plasma. \textbf{All 2D spatial snapshots are taken at a fixed cut at y=0}.
\label{fig:fig1}}
\end{figure*}

The model employs a gravitationally stratified, two-layer isothermal atmosphere: a chromosphere at $T= 10^{4}$ K and a corona at  $T= 1$ MK. The initial magnetic configuration ,as shown by Fig \ref{fig:fig1} \textbf{ a-b}, employs a modified Titov \& Démoulin \citep{1999A&A...351..707T} framework, incorporating three distinct components:

\noindent (1). The primary component ($B_{I}$) arises from a pre-existing flux rope  with major radius $R$, carrying a uniformly distributed axial current $I$ within a minor radius $a$. 

\noindent (2). The secondary component ($B_{q}$) originates from a pair of subphotospheric magnetic sources of opposite-polarity charges ($\pm q_{1}$),  separated by distance $L$ at depth $z=-d_{1}$. These charges are symmetrically positioned about the flux rope axis, mimicking active region topology. 

\noindent (3).  The tertiary component ($B_{t}$) comprises a buried dipole ($q_{2}$) positioned at $z=-d_{2}$ in the $xy$- plane, providing additional magnetic confinement. 

\noindent This configuration produces a  height-dependent magnetic field that decays as   $z^{-2}$, consistent with typical solar coronal conditions. 

\section{Results of the Simulation } \label{sec:result}

\subsection{the Global Evolution}

Figure \ref{fig:fig1} \textbf{c-f} displays the spatial distributions of key physical parameters on the xz-cross-section (at y = 0) at t = 133.09 s:  logarithmic density, temperature,  current density, and  velocity divergence. During the ascent of the filament, the magnetic fields overlaying it were \textbf{stretched up} to  create a vertical current sheet  (marked as CS in Fig.~\ref{fig:fig1} \textbf{c})  and trigger magnetic reconnection. A typical occurrence associated with this reconnection is the flare process observed, as indicated in Fig.~\ref{fig:fig1} \textbf{c} (e.g., \cite{2005ApJ...622.1251L}).

Along with the reconnection process continually converts magnetic energy into plasma kinetic and thermal energy, the CS is characterized by \textbf{parameters of the plasma} such as plasma density, temperature, and current density. As shown in Fig.~\ref{fig:fig1} \textbf{c}, the plasma density in the CS increases significantly up to twice the background value at its central region. The CS has a sharp temperature in the center reaching 14.4 MK, and a little lower value of around 14.2 MK at its top and bottom ends as shown in Fig.~\ref{fig:fig1} \textbf{d}, 
\textemdash ~an order of magnitude higher than the ambient corona temperature (1 MK). This temperature distribution aligns well with observations in \cite{2005ApJ...622.1251L}. Furthermore, the current density in the CS exhibits spatial inhomogeneity, peaking at the center and rapidly dissipating from the inside out within the ``Separatrix", as shown in Fig.~\ref{fig:fig1} \textbf{e}.

To explore the dynamic activity during the reconnection, we perform a divergence analysis of the velocity field, as described by Equation \ref{mass_conservation}, where velocity divergence effectively captures spatial density perturbations. As evidenced in Fig.~\ref{fig:fig1} \textbf{f}, we have newly found multiple outward-propagating wave fronts appearing as ripple structures originating from the ``Separatrix" near the ends of the CS. Understanding the excitation of such waves and their properties is crucial for a deeper comprehension of energy conversion mechanisms and plasma dynamics during the magnetic reconnection. A comprehensive examination of these waves will be presented in subsequent sections.

\subsection{the Characteristics of These Ripple-like Waves}

\begin{figure*}[ht!]
\begin{interactive}{animation}{./figures/Fig2.mp4}
\plotone{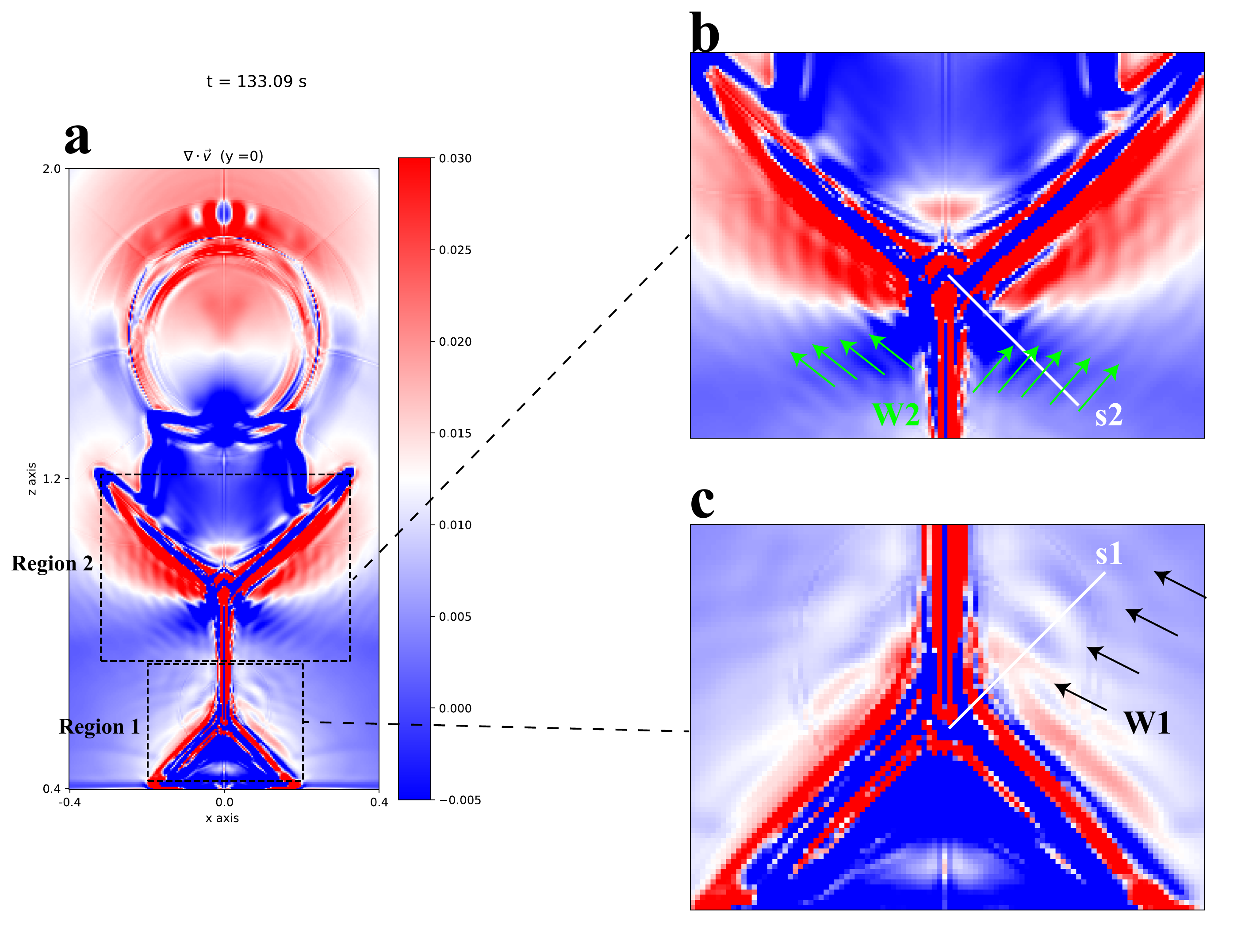}
\end{interactive}
\caption{Ripple-like waves shown by velocity divergence distribution as displayed in Panel (\textbf{a}).  Panel (\textbf{b}) is the close-up of Region 2 of Panel (\textbf{a}) to show the group wave ``W2" (green arrows) near the top end of the CS, propagating outside near the CME. Panel (\textbf{c}) is the close-up of Region 1 to display the group wave ``W1" (black arrows) near the bottom end of the CS. ``S1" in panel \textbf{c} and ``S2" in panel \textbf{b} are used to indicate the positions for the wave periodicity analysis in Fig. \ref{fig:fig3}.  An animated version from t = 132.31s to 239.71 s  of this figure is available. \label{fig:fig2}}
\end{figure*}

We calculate the velocity divergence distribution at each moment throughout the magnetic reconnection process. The observed  ripple-like waves initially manifested at both ends of the CS about 100 seconds after the onset of magnetic reconnection, becoming distinctly visible at $t = 133.09$~seconds, as illustrated in Fig ~\ref{fig:fig2}. In the detailed views provided in Fig.~\ref{fig:fig2}\textbf{b}-\textbf{c}, the wave in the Region 2 of Fig ~\ref{fig:fig2}\textbf{a} exhibits a more symmetrical feature to the CS structure compared to that in Region 1. These waves are henceforth referred to as ``W1" and ``W2" for brevity in the following descriptions.

Both ``W1" and ``W2" are simultaneously  generated by the same magnetic process within the CS but are related to opposite outflows. As shown in Fig.~\ref{fig:fig2}\textbf{b-c}, ``W2" propagates beneath the base of the CME, while ``W1" propagates above the ascending post-flare loops. In the 2D cross-section of Fig.~\ref{fig:fig2}\textbf{b-c}, ``W2" travels with a wider angular span compared to ``W1". The attenuation intensity is noticeable during their propagation, indicated by the arrows in Fig.~\ref{fig:fig2}\textbf{b-c}. Based on these simulation results, ``W2" is anticipated to be observed following the propagation of a CME, whereas ``W1" may be detected in the lower flare process.

\begin{figure*}[ht!]
\plotone{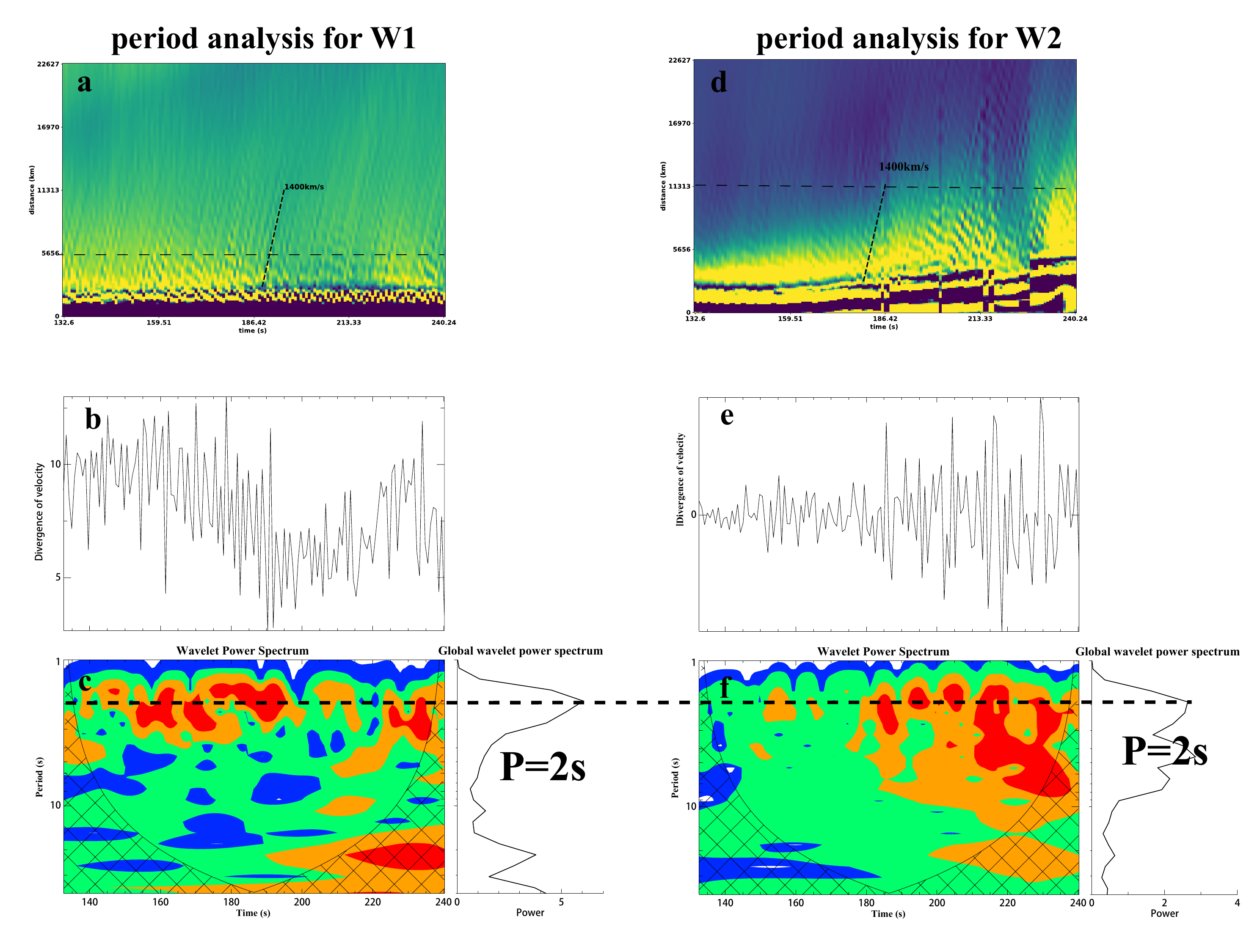}
\caption{Feature analysis of two group waves (``W1" and ``W2"). Panel \textbf{a-c} presents the period analysis for ``W1": \textbf{(a)} spatiotemporal distribution of slice ``S1"; \textbf{(b)} horizontal dashed line within \textbf{(a)} ; \textbf{(c)} wavelet analysis plot of \textbf{(a2)}. Panel \textbf{(d-f)} presents the corresponding analysis for ``W2". 
\label{fig:fig3}}
\end{figure*}

The periods and velocities of ``W1-W2" can be discerned from time-slice charts along their paths. Illustrated in Fig. \ref{fig:fig2}\textbf{c}, a slice ``s1" is taken across ``W1". The corresponding time-slice chart in Fig. \ref{fig:fig3}\textbf{a} reveals the periodic nature ``W1", with a velocity  reaching about 1400~km/s through linear fitting. According to the reference \cite{2012ApJ...753...52L}, ``W1" aligns with the typical \textbf{QFP waves} observed in the corona. Additionally, an analysis of values at the horizontal position y = 5656 km in Fig. \ref{fig:fig3}\textbf{a} estimates the period of ``W1"  using wavelet transform analysis, revealing a dominant period of 2 seconds in Fig. \ref{fig:fig3}\textbf{b-c}. Employing a similar analysis on ``W2", it is determined to propagated at a speed of $\approx$ 1400 km/s with a period of 2 seconds. The comparable velocities and periods of ``W1-W2" indicate that they were likely triggered by a similar mechanism within a shared process.

\subsection{the Origin of QFP waves}
\begin{figure*}[ht!]
\plotone{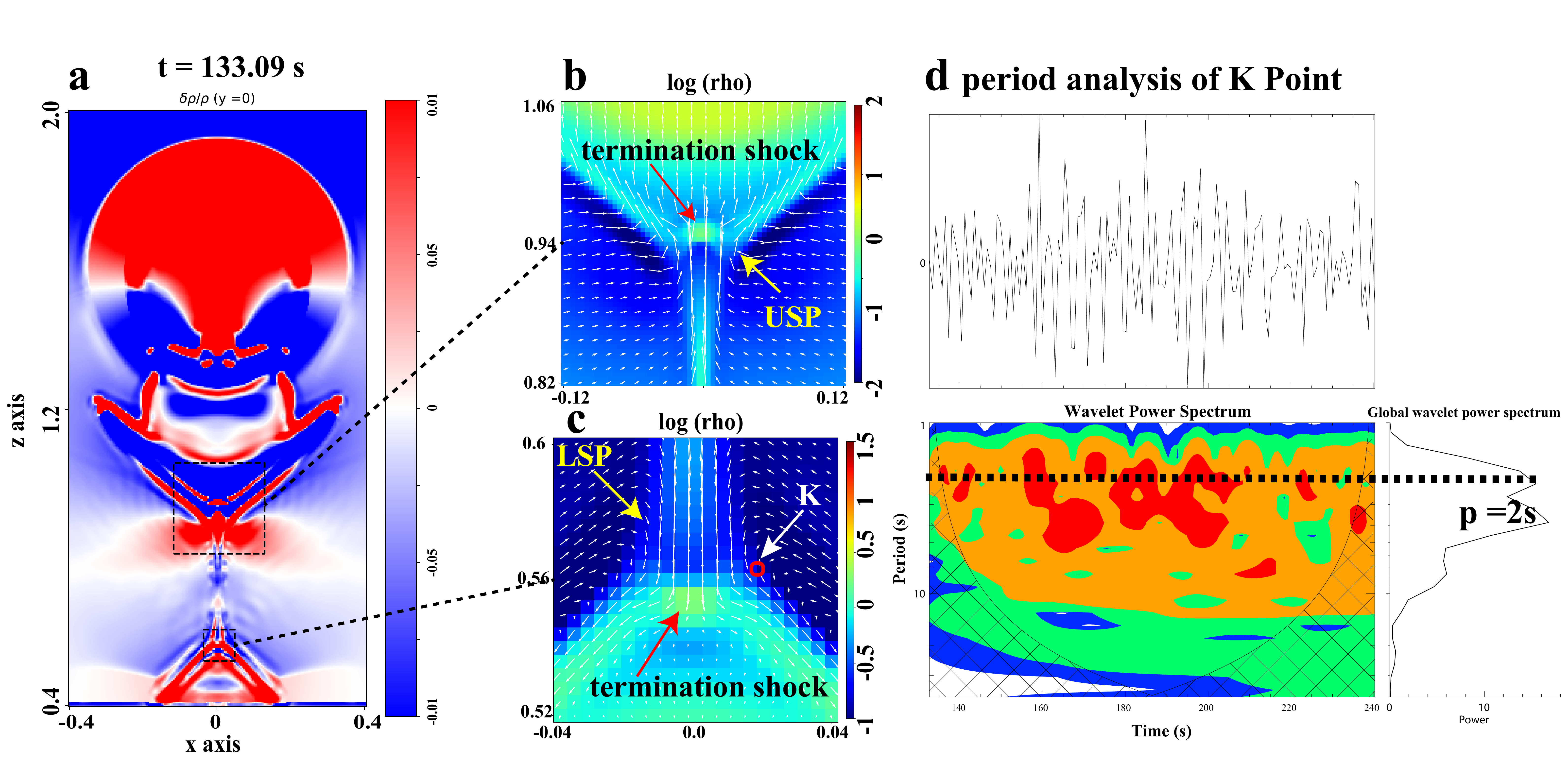}
\caption{ Distributions of density and velocities revealing the locations of fast wave excitations. \textbf{ b-c} display specific regions within the upper and lower dashed boxes in Panel \textbf{a}, where velocity fields are illustrated with white arrows, indicating the directions and magnitudes of plasma flow.  ``USP" and ``LSP" denote the upper and lower sections of the ``Separatrix" structure, respectively.  The circled region region ``K" is situated within the lower ``Separatrix" and exhibits a density oscillation of 2 s, as depicted in Panel \textbf(d).
\label{fig:fig4}}
\end{figure*}

To systematically investigate the excitation mechanism of ``W1 waves " and ``W2 waves", we need to first identify their oscillation sources and subsequently elucidate the physical processes driving these source oscillations. By tracking the temporal evolution of these waves (see the supplementary movie), we determine that these waves originate from oscillations of the ``Separatrix" structure. To verify this, we analyze oscillations at a point labeled ``K" (as shown in Fig. \ref{fig:fig4}\textbf{c}), which lies on the ``Separatrice" structure. Wavelet analysis reveals that the oscillation period at the location ``K" aligns precisely with that of the ``W1 waves", thereby confirming the ``Separatrix" as the oscillation source.

Examining the distributions of density structures and velocity fields, particularly, within the top and bottom regions of the reconnection CS, provides insights into the excitation mechanism of ``Separatrix" oscillation. The density distributions in Fig. \ref{fig:fig4} \textbf{ b-c} reveal two regions of density enhancement at the top and bottom ends of the CS, previously identified as a termination shock structure in prior  studies (eg., \cite{2017ApJ...848..102T}). Notably, following the termination shock regions (the red arrows), the high-speed reconnection outflows noticeably decelerate. The reconnection flow slows from around 720~km/s to approximately 50~km/s. Therefore, this termination shock acts as a physical barrier, and subsequently leads to a shift in the directions of the reconnection outflows  and then the interactions of reconnection outflow with ``LSP" (lower part of ``Separatrix " structure), indicated in Fig. \ref{fig:fig4}\textbf{c}.

Applying the same analytical methodology the ``W2 waves" (Fig. \ref{fig:fig4}b) revealed an identical generation mechanism: ``W2 waves" similarly originate from the impact of the reconnection outflow on a termination shock, subsequently lateral flows inducing ``USP" (upper part of the ``Separatrix" structure) oscillations. It is crucial to emphasize that ``USP" and ``LSP" belong to the same ``Separatrix" structure formed by a unified magnetic reconnection process. This topological connection implies that different segments of the ``Separatrix" structure share identical physcial properties.

 As a direct consequence, despite spatially distinct excitation sites for ``W1 waves" and ``W2 waves ", both wave systems share: (1) the same magnetic reconnection context, (2) identical excitation mechanism, and (3) consequently exhibit consistent propagation characteristics ($\sim 1400$ km/s propagation speed, and 2 s wave period).  This spatiotemporal coherence provides compelling evidence for the integral nature of the magnetic reconnection process governing the entire system.
 
\section{Discussions and Summary} \label{sec:sum}

\begin{figure*}[ht!]
\plotone{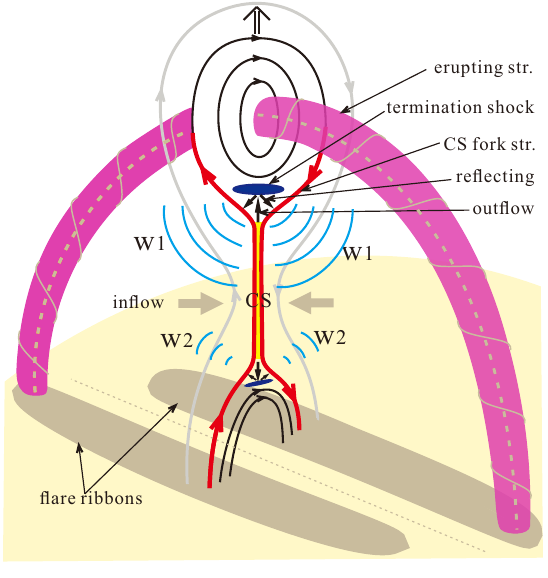}
\caption{Schematic diagram illustrating the generation of \textbf{QFP waves}. During a solar eruption, magnetic reconnection takes place within a vertical CS, propelling the erupting structure as shown by the pink twisted topology. The vertical CS is highlighted by the thick yellow lines and exhibits two fork structures at its opposing ends. Inflow during reconnection is indicated by the bold grey arrows, while reconnection outflows are represented by two thick vertical arrows at each ends. These outflows strike the termination shock structures, as indicated by the navy blue ellipses, leading to reflection. The sky blue ripple structures denote ``W1" and ``W2".\label{fig:fig5}}
\end{figure*}

The entire process can be illustrated in the schematic diagram presented in Figure \ref{fig:fig5}. The high-speed reconnection outflow (depicted by the vertical thick arrows) encounters termination shock structures (highlighted in the navy blue ellipse regions), which then rebound to impact and distort the localized sections of the CS fork wall (visible in red at both ends of the CS). This interaction induces a consistent oscillation at a specific frequency, akin to a tuning fork effects. Consequently, \textbf{QFP waves} are excited, as evidenced by the blue ripple patterns in Figure \ref{fig:fig5}. This sequence establishes a coherent causal link from magnetic reconnection to the generation of fast waves, with the termination shock playing a crucial role in facilitating the energy transfer between the reconnection outflow and the coronal waves.

Previous studies frequently demonstrate that quasi-periodic pulsations (QPPs) observed in individual events manifest a broad range of periodicity spanning from milliseconds to minutes. These extensive periodicity distributions may imply the existence of diverse physical mechanisms at play. This research unveils a new mechanism for the production of \textbf{QFP waves}, revealing the intricate nature of 3D physical process and provide new insights beyond flare model.\textbf{Moreover, quasi-periodic pulsations with the oscillation periods about those obtained in the modelling have been detected with an indirect spatial resolution in the radio emission by \cite{2018ApJ...861...33K}.}

In conclusion, by utilizing a 3D numerical model, we unveiled a novel manifestation of \textbf{QFP waves} emerging near both terminations of the vertical CS during flaring reconnection processes. These fast waves, labeled as ``W1" and ``W2" in our study, originate from the fork structure located at two ends of the vertical reconnection CS, respectively. Both ``W1"  and ``W2" exhibit similar propagation speeds of approximately 1400~km/s and share a period of 2 seconds, manifesting predominantly within the lower corona.  
``W1" demonstrates a broader angular width, whereas ``W2" presents a narrower profile. Our analysis suggests that rapid coronal waves were excited by a type of tuning fork effect occurring within the fork topology at both ends of the reconnection CS. Subsequent investigations will focus on clarifying the key role played by these fast-moving coronal waves. All in all, by leveraging the critical structure of the ``Separatrix" (or tuning fork wall), this study firstly organically connects the generation mechanisms of two groups of fast-mode coronal waves during eruptive processes. This provides a novel theoretical framework for gaining an in-depth understanding of wave phenomena driven by magnetic reconnection.

In comparing our findings with observations, we anticipate that the fast coronal wave stemming from the lower end of the CS should be detectable in proximity to post-flare loops, as already confirmed through EUV observations (e.g.,\cite{2011ApJ...736L..13L}). Conversely, the fast coronal wave originating from the upper end of the CS is anticipated to be observed in conjunction with propagating  eruptive coronal mass ejections (CMEs),  a phenomenon awaiting observation validation.

It is important to note a key limitation of our present analysis: the characteristics of these QFP waves are analysed in the vertical plane passing through the center of the simulation domain. In a fully three-dimensional corona, these waves are subject to additional effects not captured here, such as refraction induced by the 3D magnetic field structure (e.g., \cite{2025ApJ...990....1S}).  A comprehensive investigation of these 3D effects will be a crucial and natural extension of this work in the future. Moreover, this study serves as a foundational case study and a key outstanding question remains regarding what physical parameters primarily determine the resulting oscillation period. A systematic parametric study to quantify the range of the periods is therefore essential for future work.

\vspace{2em}
\noindent
The work is supported by the National Key R\&D Program of China (No. 2022YFF0503800),  the Strategic Priority Research Program of the Chinese Academy of Sciences (No. XDB0560102), and the National Natural Science Foundation of China (Nos. 12503064, 12273061, 12403067,12303062). We also acknowledge Sichuan Normal University Astrophysical Laboratory Supercomputer for providing the computational resources.

\bibliography{sample701}{}
\bibliographystyle{aasjournalv7}

\end{document}